%% file: main.tex
\documentclass[10pt,twocolumn,letterpaper]{article}

\usepackage[pagenumbers]{cvpr} % To force page numbers, e.g. for an arXiv version

\input{preamble}
\definecolor{cvprblue}{rgb}{0.21,0.49,0.74}
\usepackage[pagebackref,breaklinks,colorlinks,allcolors=cvprblue]{hyperref}
\usepackage{multirow}
\usepackage{float}
\usepackage{xcolor,colortbl}
\usepackage{amsmath,amssymb,amsthm}

\definecolor{lightred}{RGB}{255, 204, 204}
\definecolor{strongred}{RGB}{255, 102, 102}

\def\paperID{13419} % *** Enter the Paper ID here
\def\confName{CVPR}
\def\confYear{2026}

\title{Variable Basis Mapping for Real-Time Volumetric Visualization}

\author{
Qibiao Li\thanks{Equal contribution} \quad 
Yuxuan Wang\footnotemark[1] \quad 
Youcheng Cai\thanks{Corresponding author} \quad 
Huangsheng Du \quad 
Ligang Liu\\[5pt]
University of Science and Technology of China\\
{\tt\small caiyoucheng@ustc.edu.cn}
}

\begin{document}
\maketitle
\input{sec/0_abstract}    
\input{sec/1_intro}

\input{sec/2_rel}
\input{sec/3_pre}
\input{sec/4_meth}

\input{sec/5_exp}

{
    \small
    \bibliographystyle{ieeenat_fullname}
    \bibliography{main}
}

% \begin{figure*}
%   \centering
%   \begin{subfigure}{0.68\linewidth}
%     \fbox{\rule{0pt}{2in} \rule{.9\linewidth}{0pt}}
%     \caption{An example of a subfigure.}
%     \label{fig:short-a}
%   \end{subfigure}
%   \hfill
%   \begin{subfigure}{0.28\linewidth}
%     \fbox{\rule{0pt}{2in} \rule{.9\linewidth}{0pt}}
%     \caption{Another example of a subfigure.}
%     \label{fig:short-b}
%   \end{subfigure}
%   \caption{Example of a short caption, which should be centered.}
%   \label{fig:short}
% \end{figure*}


\end{document}

\maketitle
\input{sec/X_supp}
{
    \small
    \bibliographystyle{ieeenat_fullname}
    \bibliography{supp}
}
% \begin{figure*}
%   \centering
%   \begin{subfigure}{0.68\linewidth}
%     \fbox{\rule{0pt}{2in} \rule{.9\linewidth}{0pt}}
%     \caption{An example of a subfigure.}
%     \label{fig:short-a}
%   \end{subfigure}
%   \hfill
%   \begin{subfigure}{0.28\linewidth}
%     \fbox{\rule{0pt}{2in} \rule{.9\linewidth}{0pt}}
%     \caption{Another example of a subfigure.}
%     \label{fig:short-b}
%   \end{subfigure}
%   \caption{Example of a short caption, which should be centered.}
%   \label{fig:short}
% \end{figure*}


%% file: sec/0_abstract.tex
\begin{abstract}
Real-time visualization of large-scale volumetric data remains challenging, as direct volume rendering and voxel-based methods suffer from prohibitively high computational cost. We propose \textbf{Variable Basis Mapping (VBM)}, a framework that transforms volumetric fields into 3D Gaussian Splatting (3DGS) representations through wavelet-domain analysis. First, we precompute a compact Wavelet-to-Gaussian Transition Bank that provides optimal Gaussian surrogates for canonical wavelet atoms across multiple scales. Second, we perform analytical Gaussian construction that maps discrete wavelet coefficients directly to 3DGS parameters using a closed-form, mathematically principled rule. Finally, a lightweight image-space fine-tuning stage further refines the representation to improve rendering fidelity. Experiments on diverse datasets demonstrate that VBM significantly accelerates convergence and enhances rendering quality, enabling real-time volumetric visualization. The code will be publicly released upon acceptance.
\end{abstract}

%% file: sec/1_intro.tex
\section{Introduction}
\label{sec:intro}
Volumetric data are widely used across science and engineering, ranging from medical imaging (CT, MRI)~\cite{klacansky2017openscivis,niedermayr2024novel} to simulations in fluid dynamics~\cite{Jakob20,li2008public}, climate~\cite{athawale2024uncertainty}, and planetary-scale systems~\cite{li2024paramsdrag}. Volume visualization is crucial for revealing patterns in such data and for providing users with an intuitive means of complex structures.

As the resolution and scale of volumetric data increase, traditional visualization methods, such as direct volume rendering~\cite{dvr}, become computationally expensive and slow, hindering real-time interaction and flexible data exploration. Recently, 3D Gaussian Splatting (3DGS)~\cite{kerbl3Dgaussians} has emerged as a powerful technique for representing and rendering complex scenes. Unlike volumetric ray marching~\cite{weiss2021differentiable,mildenhall2021nerf} or voxel-based methods~\cite{yu2021plenoctrees,fridovich2022plenoxels,SunSC22,chen2022tensorf,muller2022instant,fridovich2023k}, 3DGS achieves real-time photorealistic rendering using explicit, anisotropic 3D Gaussian primitives and analytical rasterization. From a functional perspective, this representation can be viewed as a projection of the radiance field—a mapping from $\mathbb{R}^3$ to $\mathbb{R}^4$—onto a function space spanned by Gaussian basis functions. The optimal Gaussian approximation is obtained through an image-driven optimization, in which both initialization and the optimization trajectory determine the final quality and convergence speed.

\begin{figure}[t]
    \centering
    \includegraphics[width=1\linewidth]{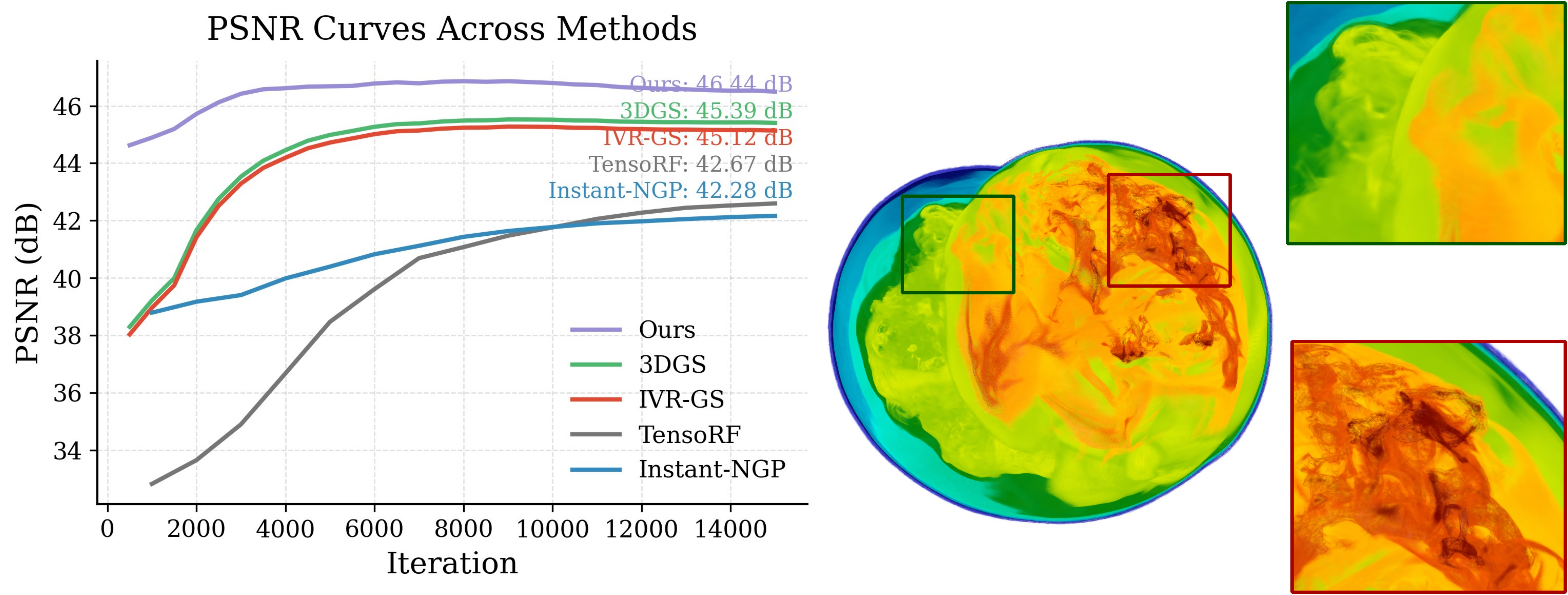}
    \caption{Left: Our excellent PSNR curve demonstrating the performance of the \emph{VBM} framework on Supernova data. Right: The VBM-based Gaussian model effectively captures fine details.}
    \label{fig:teaser-figure}
\end{figure}

In novel view synthesis, 3DGS typically relies on structure-from-motion (SfM)~\cite{schoenberger2016sfm} for initialization. For volumetric visualization, iVR-GS~\cite{tang2025ivrgs} adopts randomly initialized Gaussian primitives, whereas VEG~\cite{dyken2025volumeencodinggaussians} employs a heuristic strategy in which each initial Gaussian is defined as a sphere whose size is proportional to the local point density. Sewell et al.~\cite{Sewell2024HighqualityAO} seek to enhance initialization by sampling points from isosurfaces to position the Gaussians more accurately, although isosurface extraction can be computationally expensive. Nevertheless, these approaches still suffer from two major limitations: (1) they are heuristic in nature and fail to exploit the rich structural and statistical information inherent in volumetric data; and (2) the construction of Gaussians relies heavily on computationally intensive image-space optimization, leading to high computational cost and slow convergence. 

To overcome these limitations, we draw inspiration from the spatial–frequency decomposition of wavelets and leverage their multiresolution analysis of volumetric data to obtain high-quality initializations for 3DGS, thereby enabling stable and efficient optimization. However, employing wavelet analysis introduces two significant challenges: (1) there exists a substantial representational gap between wavelet and Gaussian bases, which makes direct conversion nontrivial; and (2) a naïve implementation would incur cubic computational complexity in Wavelet-to-Gaussian transitions, resulting in prohibitive computational cost.

To address these issues, we propose a \textbf{Variable-Basis Mapping} (VBM) framework for efficient volumetric rendering in the wavelet domain. First, we precompute a compact \textbf{Wavelet-to-Gaussian Transition Bank}. For each wavelet kernel $\psi_{j,\mathbf{k}}(x)$ at frequency level $j$ and spatial position $\mathbf{k}$, we obtain its optimal Gaussian approximation through a statistical estimation process. Although direct computation would require $J \times G^3$ individual transitions for a $J$-level wavelet transform, the translation covariance of wavelet bases allows us to compute only the canonical kernel $\psi_{j,\mathbf{k}=(0,0,0)}$ for each level, where $G$ denotes the volumetric data resolution. The remaining kernels can be derived analytically through translation and scaling, as guaranteed by the group representation theorem~\cite{ali2000coherent,antoine1999wavelets}. Second, we propose an \textbf{analytical Gaussian construction} strategy that decomposes the volumetric field using the discrete wavelet transform (DWT) \cite{dip,mallat1999wavelet}, representing the volume as a linear combination of wavelet basis functions and their corresponding coefficients. By applying the precomputed Wavelet-to-Gaussian transitions, each wavelet basis is mapped analytically to a Gaussian basis, and the associated wavelet coefficients are systematically translated into the corresponding 3DGS parameters according to a principled mapping rule. Finally, an \textbf{image-space Gaussian fine-tuning} strategy refines the Gaussian parameters, ensuring visual consistency and convergence. Experimental results demonstrate that our method not only improves training efficiency but also achieves higher rendering quality than state-of-the-art approaches.

Our main contributions can be summarized as follows:
\begin{itemize}
\item To the best of our knowledge, \textbf{Variable Basis Mapping} is the first framework that establishes a direct, mathematical bridge between volumetric analysis and 3D Gaussian Splatting, laying a new foundation for efficient field-to-primitive conversion.
\item We develop a principled kernel transition mechanism, named \textbf{Wavelet-to-Gaussian Transition Bank}, theoretically grounded in translation consistency, which enables efficient and reusable derivation of Gaussian bases.
\item We introduce a \textbf{Paradigm Shift for Real-Time Volume Visualization}, that transforms traditional volumetric visualization into 3D Gaussian Splatting format. This “baking” process preserves volumetric semantics while inheriting 3DGS’s real-time rendering efficiency.
\end{itemize}

\begin{figure*}[h]
     \centering
     \includegraphics[width=\textwidth]{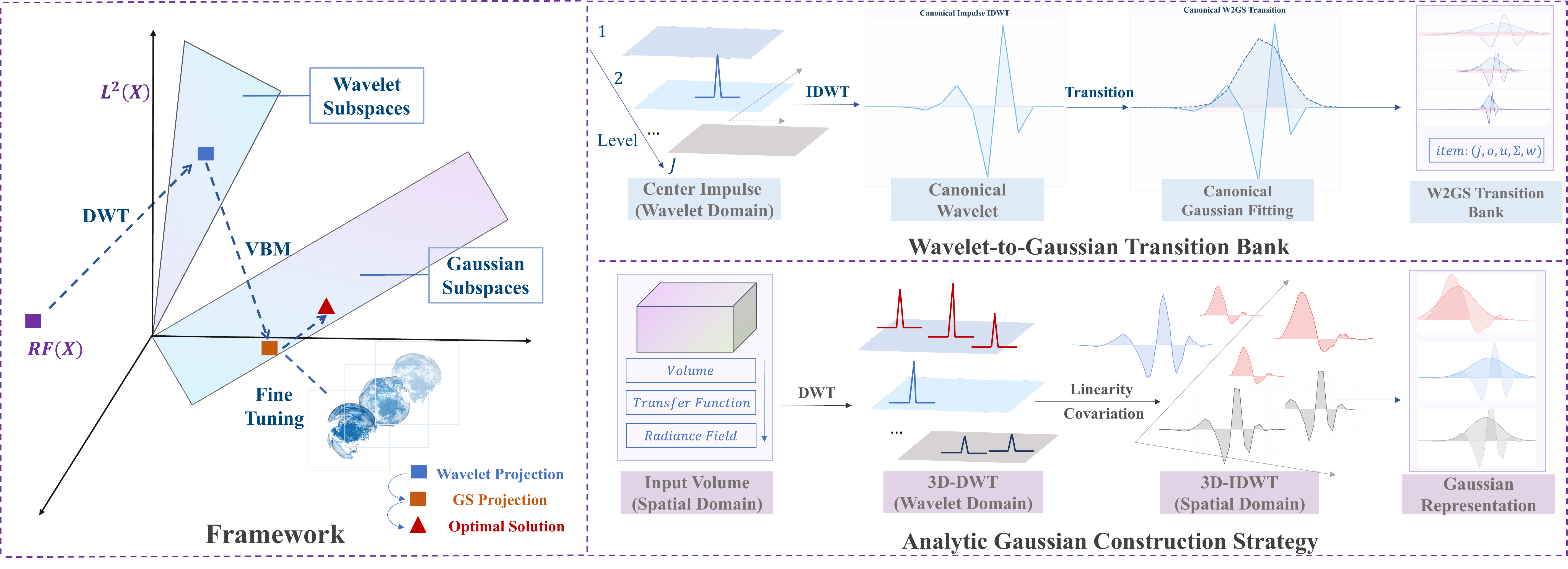}
    \caption{Overview of the \emph{VBM} framework. A volumetric radiance field is projected onto the wavelet subspace using discrete wavelet transform (DWT), and then mapped to the function subspace spanned by Gaussian primitives. This is followed by efficient image-space fine-tuning to obtain the optimal parameters. The process is supported by the Wavelet-to-Gaussian Transition Bank, built from canonical wavelets and Gaussians, and an analytic Gaussian construction strategy leveraging the linearity and translation consistency of the IDWT.}
    \label{fig:method}
\end{figure*}

%% file: sec/2_rel.tex
\section{Related Work}
\textbf{Volumetric Representations and Rendering.}
Volumetric data are central to scientific visualization, computer graphics, and medical imaging. Traditional Direct Volume Rendering~\cite{dvr} integrates radiance along viewing rays for physically accurate visualization but is computationally expensive due to dense sampling requirements.

To improve scalability, hierarchical data structures such as Sparse Voxel Octrees~\cite{laine2010efficient, outofcore-octree-1, outofcore-octree-2}, multi-resolution pyramids~\cite{mipmap, ECNR}, and bricked grids~\cite{brick1, brick2} enable adaptive resolution and memory-efficient rendering.

Rendering efficiency has also benefited from more effective sampling strategies. Importance sampling~\cite{viola2004importance, kraft2020adaptive, li2023nerfacc, kurz2022adanerf} focuses computation on radiometrically or perceptually salient regions, while learned selective rendering~\cite{makeff, FoVolNet} reconstructs full images from sparsely rendered content.

\textbf{Neural Scene Representations.}
Recent advances in neural radiance field, such as NeRF~\cite{mildenhall2021nerf}, Plenoxel~\cite{fridovich2022plenoxels}, TensoRF~\cite{chen2022tensorf}, InstantNGP~\cite{muller2022instant}, and K-Planes~\cite{fridovich2023k}, have enabled compact and high-fidelity representations for novel view synthesis. Building upon these implicit approaches, 3D Gaussian Splatting (3DGS)~\cite{kerbl3Dgaussians} represents a scene using anisotropic Gaussian primitives that are both differentiable and efficient to render, bridging volumetric and explicit representations. 
Extensions that incorporate Gaussians into mesh-based framework, such as SuGaR~\cite{guedon2023sugar}, GaussianMesh~\cite{MeshGaussian2024}, GaMeS~\cite{waczynska2024games}, and MeshGS~\cite{choi2024meshgs}, enhance surface fidelity by aligning or attaching Gaussian distributions to mesh surfaces or vertices, leveraging local geometry for improved accuracy.

For volumetric modeling, IVR-GS~\cite{tang2025ivrgs} initializes Gaussians randomly and optimizes them entirely in image space, while VEG~\cite{dyken2025volumeencodinggaussians} heuristically distributes Gaussian scales according to local density. Sewell et al.~\cite{Sewell2024HighqualityAO} further employ isosurface-based point clouds for high-quality ambient occlusion. However, these methods remain largely heuristic and fail to leverage the inherent structural and statistical regularities of volumetric data. In contrast, our approach establishes a principled mapping from volumetric field to Gaussian primitives, within a wavelet-informed framework.

\textbf{Wavelet Analysis in Volumetric Data Processing.}
Wavelet analysis~\cite{daubechies1992ten,mallat1999wavelet} offers a multiresolution representation with joint spatial–frequency localization, outperforming global transforms such as the Fourier~\cite{stein2011fourier}. Wavelets have been widely applied to volumetric compression~\cite{ihm1999wavelet,kim1999efficient,bruylants2015wavelet}, denoising~\cite{chen2004breast,jerhotova2011biomedical,chervyakov2020analysis}, and feature extraction~\cite{prochazka2011three,ghasemzadeh20183d}.

Beyond volumetric settings, wavelets have demonstrated broad utility in image processing~\cite{dip,imagecodewavelet}, time-series analysis~\cite{rhif2019wavelet,waveletsignalanalysis}, and modern deep learning pipelines~\cite{zhou2024udiff,zhang2023flight}, underscoring their versatility for hierarchical, frequency-aware signal modeling.

%% file: sec/3_pre.tex
\section{Preliminaries}

\textbf{Transfer Function.} 
Scientific visualization maps spatial or spatio-temporal data into perceptually meaningful visual representations. 
Given a spatial domain $D \subset \mathbb{R}^d$, the dataset is modeled as a field $F : D \to \mathbb{R}^r$, where $r$ denotes the value dimension. 
A \emph{transfer function} $TF: \mathbb{R}^r \to \mathbb{R}^n$ transforms physical quantities into visual attributes of a radiance field, typically producing an RGBA vector $(\mathbf{c}, \alpha)$.
In this paper, we discretize $D$ into a uniform grid of resolution $G^3$, yielding 
$F \in \mathbb{R}^{G \times G \times G \times r}$. 
Applying the transfer function produces the radiance volume:
$
RF = TF(F) \in \mathbb{R}^{G \times G \times G \times 4},
$
yielding the volume rendered downstream.

\textbf{3D Wavelet Transform.} 
The wavelet transform provides joint spatial--frequency localization, enabling compact and structured representations of volumetric fields. Given a mother wavelet $\psi(\cdot)$ and a scaling function $\phi(\cdot)$, the dyadic atoms are defined as:
\begin{equation}
\psi_{j,\mathbf{k}}(\mathbf{x}) = 2^{3j/2}\, \psi(2^{j}\mathbf{x} - \mathbf{k}).
\end{equation}
where $\mathbf{x}$ denotes the spatial position, $j$ is the scale index, and $\mathbf{k}=(k_x, k_y, k_z)$ is the spatial translation index.

Each channel $i$ of a vector-valued radiance field $RF^{(i)}(\mathbf{x})$ can be expressed as a multiresolution expansion:
\begin{equation}
RF^{(i)}(\mathbf{x}) =
\sum_{\mathbf{k}} c^{(i)}_{0,\mathbf{k}}\phi(\mathbf{x}-\mathbf{k}) + \sum_{j=0}^{J-1}\sum_{\mathbf{k}} d^{(i)}_{j,\mathbf{k}}\psi_{j,\mathbf{k}}(\mathbf{x}),
\end{equation}
where $c^{(i)}_{0,\mathbf{k}} = \langle RF^{(i)}, \phi(\mathbf{x}-\mathbf{k}) \rangle$, $d^{(i)}_{j,\mathbf{k}} = \langle RF^{(i)}, \psi_{j,\mathbf{k}} \rangle$ are the low- and high-frequency coefficients, respectively. Here, $J$ denotes the number of decomposition levels in the wavelet transform. Accordingly, the radiance field $RF(\cdot)$ can be expressed in a compact multi-scale form using the 3D wavelet transform:
\begin{equation}
RF(\mathbf{x}) \approx
\sum_{j=0}^{J}
\sum_{\mathbf{k} \in D}
\begin{bmatrix}
\mathbf{c}_{j,\mathbf{k}} \\ \alpha_{j,\mathbf{k}}
\end{bmatrix} \, \psi_{j,\mathbf{k}}(\mathbf{x}),
\label{eq:rfdwt}
\end{equation}
where the coefficient vectors $\mathbf{c}_{j,\mathbf{k}} = \langle \mathbf{c}, \psi_{j,\mathbf{k}} \rangle$ and $\alpha_{j,\mathbf{k}} = \langle \alpha, \psi_{j,\mathbf{k}} \rangle$ encode the color and opacity  at different scales.

\textbf{3D Gaussian Splatting.} 
3DGS~\cite{kerbl3Dgaussians} represents the radiance field as a collection of $N$ anisotropic Gaussian kernels:
\begin{equation}
\mathcal{G}
= \big\{ (\boldsymbol{\mu}_i, \boldsymbol{\Sigma}_i,
\mathbf{c}_i, \alpha_i) \big\}_{i=1}^{N},
\end{equation}
where $\boldsymbol{\mu}_i \!\in\! \mathbb{R}^3$ denotes the center,
$\boldsymbol{\Sigma}_i \!\in\! \mathbb{R}^{3\times3}$ represents the covariance matrix,
$\mathbf{c}_i$ is the color, and $\alpha_i$ is the opacity.

The radiance field can be approximated by a weighted linear combination of anisotropic Gaussian primitives:
\begin{equation}
RF(\mathbf{x})
\approx
\sum_{i=1}^{N}
\begin{bmatrix}
\mathbf{c}_i \\ \alpha_i
\end{bmatrix}
\exp\!\left[
-\tfrac{1}{2}
(\mathbf{x}-\boldsymbol{\mu}_i)^{\!\top}
\boldsymbol{\Sigma}_i^{-1}
(\mathbf{x}-\boldsymbol{\mu}_i)
\right].
\end{equation}
This linear combination over localized Gaussian primitives enables an efficient and differentiable approximation of the radiance distribution, while supporting real-time rendering.

%% file: sec/4_meth.tex
\begin{figure}
     \centering
     \includegraphics[width=1.0\linewidth]{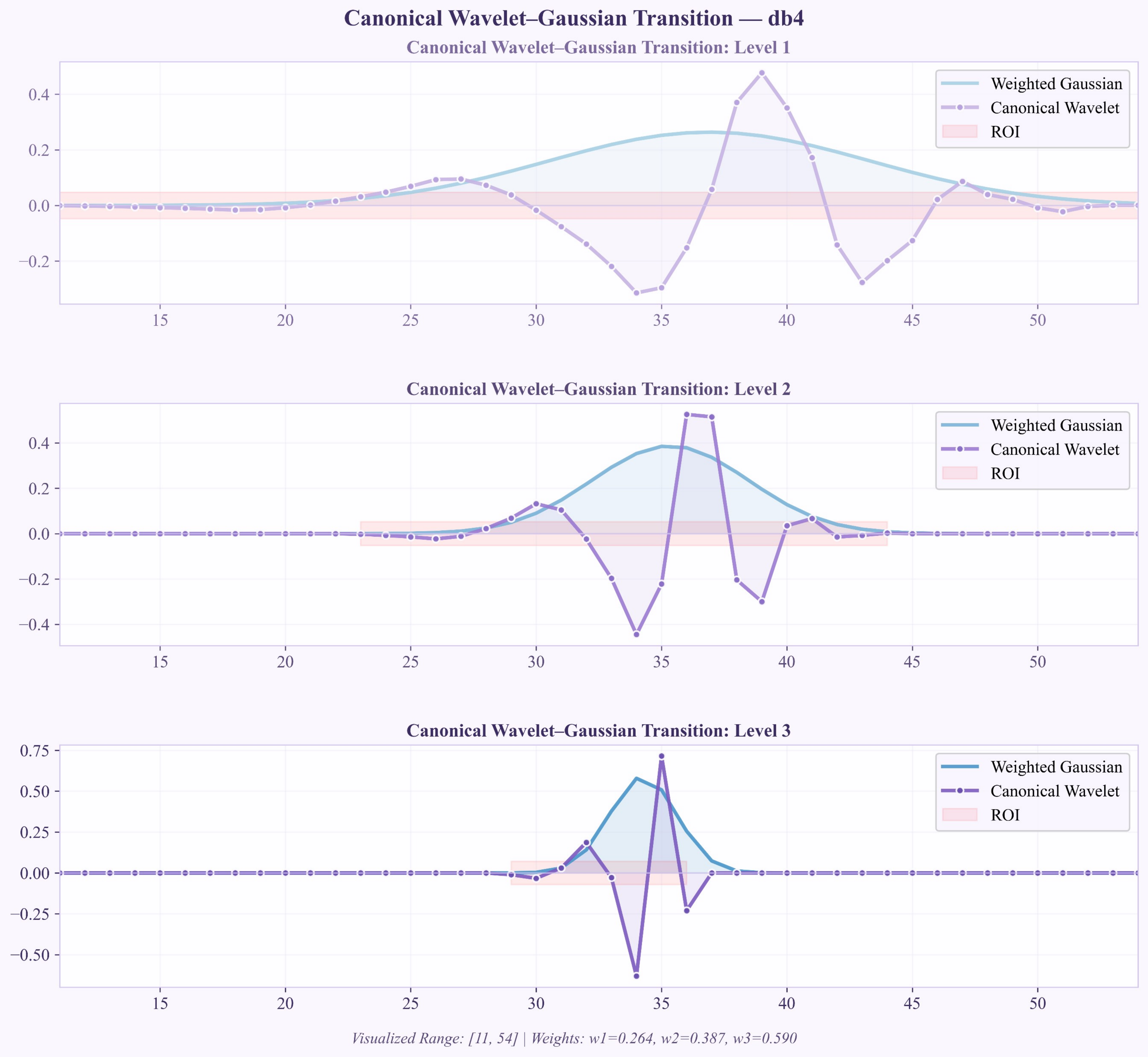}
     \caption{
        \textbf{Wavelet-to-Gaussian Transition Bank.}
         Illustration of the transition from localized wavelet kernels to Gaussian primitives across multiple scales.
     }
     \label{fig:w2gs}
\end{figure}

\section{Method}
\label{sec:meth}

\subsection{Formulation and Overview}
\label{sec:formulation}
We cast \emph{Variable Basis Mapping} as a sequence of functional mapping connecting the physical field, its visual encoding, and final primitive representation. Given a physical field $F$, we first apply a transfer function $TF$ to obtain a radiance field $RF$. The radiance field is then decomposed into its multi-scale wavelet representation $\Phi$, which is subsequently converted into a set of 3D Gaussian primitives $\mathcal{G}$. The overall mapping is summarized as:
$ F \xrightarrow{TF} RF \xrightarrow{} \Phi \xrightarrow{} \mathcal{G}.$

We have established an efficient correspondence between wavelets and their optimal Gaussian counterparts:
\begin{equation}
    \psi_{j,\mathbf{k}}(\mathbf{x}) 
    \sim g_{j,\mathbf{k}}(\mathbf{x}).
    \label{eq:w2g_translation}
\end{equation}
Substituting Eq.~\eqref{eq:w2g_translation} into Eq.~\eqref{eq:rfdwt}, we obtain:
\begin{equation}
    RF(\mathbf{x}) \approx
    \sum_{j=0}^{J-1} \sum_{\mathbf{k}} 
     \begin{bmatrix}
        \mathbf{c}_{j,\mathbf{k}},\, 
        \alpha_{j,\mathbf{k}}
    \end{bmatrix}^\top g_{j,\mathbf{k}}(\mathbf{x}),
    \label{eq:volume2gs}
\end{equation}
where $\begin{bmatrix}\mathbf{c}_{j,\mathbf{k}},\,\alpha_{j,\mathbf{k}}\end{bmatrix}^{\top}
= (R,G,B,\alpha)^{\!T}$ denotes the color and opacity coefficients, jointly determined by the wavelet subband intensity and an analytically derived parameter mapping rule. This formulation directly yields the Gaussian representation $\mathcal{G}$ of the volumetric field.

\textbf{Overview.} Our goal is to exploit the information of volume to construct Gaussian primitives directly, thereby enabling efficient real-time rendering and optimization.
We first develop a Wavelet-to-Gaussian Transition Bank, which provides precomputed Gaussian approximations for canonical wavelet kernels (Sec.~\ref{sec:4.2}). 
Next, we apply a 3D discrete wavelet transform (DWT) to the radiance field to extract its spatial-frequency characteristics, and leveraging the transition bank, derive a Gaussian-splatting expansion via a simple lookup operation (Sec.~\ref{sec:4.3}).
Finally, a lightweight image-space fine-tuning stage refines the resulting 3DGS representation, enabling real-time volumetric visualization (Sec.~\ref{sec:4.4}). See Fig.~\ref{fig:method} for an overview of the process.

\subsection{Wavelet-to-Gaussian Transition Bank}
\label{sec:4.2}
We develop an efficient procedure to construct Gaussian surrogates for wavelet kernels, as depicted in Fig.~\ref{fig:w2gs}. Given a $J$-level wavelet transform applied to the radiance field $RF$, a straightforward approach would independently fit an optimal Gaussian to every wavelet kernel across all scales and spatial positions, leading to $J \times G^3$ fittings. However, within each subband, all wavelet kernels are identical up to translation. Thus, it is sufficient to fit a \emph{single} canonical Gaussian per subband, and obtain others via spatial translation.

Concretely, our procedure consists of four steps: (i) define a \emph{canonical wavelet kernel}, (ii) derive its \emph{translation rule} to arbitrary spatial locations, (iii) fit an optimal Gaussian to the canonical kernel, and (iv) translate the fitted parameters to generate approximations for any kernel within the same subband. This yields a compact yet expressive {Wavelet-to-Gaussian Transition Bank} that supports analytical conversion from wavelet bases to Gaussian primitives.

\textbf{Canonical Wavelet Construction}. 
We define the canonical wavelet kernel as the spatial footprint of a unit impulse in the wavelet domain. Let $\mathcal{W}$ denote the 3D discrete wavelet transform (DWT) with $J$ decomposition levels. At each level $j$, subbands are indexed by:
\begin{equation}
\mathbf{o} = (o_x, o_y, o_z) \in \{\texttt{L}, \texttt{H}\}^3,
\end{equation}
where $o_d \in \{\texttt{L}, \texttt{H}\}$ indicates low- or high-pass filtering along axis $d \in \{x,y,z\}$.  
For instance, $\mathbf{o}=(\texttt{L},\texttt{L},\texttt{H})$ corresponds to low-pass filtering on $x,y$, and high-pass on $z$.

Given a decomposition level $j$ and orientation $\mathbf{o}$, we define an \emph{impulse} in the corresponding subband as:
\begin{equation}
e_{j,\mathbf{o},\mathbf{k}=(k_x,k_y,k_z)}(\boldsymbol{\kappa}) =
\begin{cases}
1, & \text{if } \boldsymbol{\kappa} = \mathbf{k} \ \text{in subband } (j,\mathbf{o}), \\[4pt]
0, & \text{otherwise},
\end{cases}
\label{eq:wavelet_impulse}
\end{equation}
where $\mathbf{k} = (k_x, k_y, k_z)$ denotes the spatial index. 

Applying the inverse wavelet transform, $\mathcal{W}^{-1}$, to this one-hot feature yields its spatial-domain response. Further, by centering the impulse at $\mathbf{k} = (0, 0, 0)$, we obtain the \emph{canonical wavelet}:
$\hat{\psi}_{j,\mathbf{o}}(\mathbf{x})$:
\begin{equation}
\big\{
\hat{\psi}_{j,\mathbf{o}}(\mathbf{x})
= \mathcal{W}^{-1}\!\left[e_{j,\mathbf{o},(0,0,0)}(\boldsymbol{\kappa}) \right]\!
\;\big|\;
j,
\mathbf{o}
\big\},
\label{eq:canonical_wavelet}
\end{equation}
where $j \in \{0,\dots,J\}$, and  $\mathbf{o} \in \{\texttt{L},\texttt{H}\}^3$. The complete set $\{\hat{\psi}_{j,\mathbf{o}}(\mathbf{x})\}$ therefore contains $8(J + 1)$ canonical wavelet kernels, each capturing a distinct scale–orientation pattern in the volumetric field.

\textbf{Translation Consistency.}
We now establish translation consistency within each subband: all wavelet kernels in a given subband are spatially translated replicas of the corresponding canonical kernel. Consider an arbitrary impulse in subband $(j,\mathbf{o})$ located at $\mathbf{k}$. Its spatial-domain response is given by:
\begin{equation}
\hat{\psi}_{j,\mathbf{o},\mathbf{k}}(\mathbf{x})
:= \mathcal{W}^{-1}\!\big[e_{j,\mathbf{o},\mathbf{k}}\big](\mathbf{x}),
\label{eq:idwt_impluse}
\end{equation}
which can be derived from the canonical response at the same scale and orientation.

To formalize this relationship, we introduce the translation operator 
$(T_{\boldsymbol{\Delta}}f)(\mathbf{x}) = f(\mathbf{x}-\boldsymbol{\Delta})$. 
Under approximately linear-phase filter banks, such as \emph{biorthogonal} wavelets~\cite{daubechies1992ten} with symmetric or periodic boundary extensions, the inverse DWT exhibits near-translation covariance on the discrete sampling grid:
\begin{equation}
\hat{\psi}_{j,\mathbf{o},\mathbf{k}}(\mathbf{x})
\;\approx\;
(T_{\,2^{j}\mathbf{k}+\boldsymbol{\delta}_{j,\mathbf{o}}}\,
\hat{\psi}_{j,\mathbf{o}})(\mathbf{x}),
\label{eq:covariation_law}
\end{equation}
where $\boldsymbol{\delta}_{j,\mathbf{o}}$ is a subpixel offset determined by the phase delay and half-sample shift of the analysis–synthesis filter pair. In practice, we use the biorthogonal 4.4 wavelet and set $\boldsymbol{\delta}_{j,\mathbf{o}} \approx (\tfrac{1}{2}, \tfrac{1}{2}, \tfrac{1}{2})$.

This property guarantees that all wavelets sharing the same scale $j$ and orientation $\mathbf{o}$ are spatially consistent translations of the canonical response $\hat{\psi}_{j,\mathbf{o}}(\mathbf{x})$. Consequently, Gaussian primitives fitted to the canonical kernel can be efficiently propagated across spatial positions through simple translation, eliminating the need for redundant parameter fitting for each wavelet instance.

\textbf{Transition Bank Construction.}
Here, we approximate the canonical wavelet kernel with a single Gaussian primitive. The aim is to match the canonical wavelet’s dominant spatial footprint with a Gaussian envelope, a procedure we term \emph{dominant-lobe estimation}.
Given the canonical wavelet kernel $\hat{\psi}_{j,\mathbf{o}}(\mathbf{x})$
we first compute a normalized magnitude distribution:
\begin{align}
&W(\mathbf{x})
=
\frac{|\hat{\psi}_{j,\mathbf{o}}(\mathbf{x})|}
     {\sum_{\mathbf{u}\in D}|\hat{\psi}_{j,\mathbf{o}}(\mathbf{u})|},
\nonumber\\[4pt]
&\mathcal{R}
=
\big\{
\mathbf{x}\in D :
W(\mathbf{x}) > \tau\,\max_{\mathbf{u}\in D} W(\mathbf{u})
\big\},
\label{eq:roi}
\end{align}
where $\tau\in(0,1)$ is the relative threshold that defines \emph{the region of interest} (ROI) corresponding to the dominant lobe. Within this ROI, we estimate the centroid and covariance of the kernel from the first- and second-order weighted moments:
\begin{align}
\mathbf{u}_{j,\mathbf{o}} &=
\frac{\sum_{\mathbf{x}\in\mathcal{R}}
      W(\mathbf{x})\,\mathbf{x}}
     {\sum_{\mathbf{x}\in\mathcal{R}} W(\mathbf{x})},
\nonumber\\[4pt]
\Sigma_{j,\mathbf{o}} &=
\frac{\sum_{\mathbf{x}\in\mathcal{R}}
      W(\mathbf{x})\,(\mathbf{x}-\mathbf{u}_{j,\mathbf{o}})(\mathbf{x}-\mathbf{u}_{j,\mathbf{o}})^{\!\top}}
     {\sum_{\mathbf{x}\in\mathcal{R}} W(\mathbf{x})}.
\label{eq:moment}
\end{align} 
These two parameters define the geometry of the canonical Gaussian approximation:
\begin{equation}
\big\{
g_{j,\mathbf{o}}(\mathbf{x})
\;=\;
\exp\!\Big(
-\tfrac{1}{2}
(\mathbf{x}-\mathbf{u}_{j,\mathbf{o}})^{\!\top}
\Sigma_{j,\mathbf{o}}^{-1}
(\mathbf{x}-\mathbf{u}_{j,\mathbf{o}})
\Big)
\;\big|\;
j,
\mathbf{o}
\big\}.
\label{eq:canonical_gaussian}
\end{equation}

\textbf{Energy Conservation.}
To preserve energy between each wavelet kernel and its Gaussian approximation, we introduce a weight vector $\mathbf{w}_{j,\mathbf{o}}\!\in\!\mathbb{R}^4$ that rescales the Gaussian envelope to match the original wavelet energy. 
The scalar weight \(w_{j,\mathbf{o}}\) is obtained minimizing the $L^2$ discrepancy between the wavelet magnitude and the scaled Gaussian:
\begin{equation}
w_{j,\mathbf{o}}
=
\arg\min_{w}
\big\|
|\hat{\psi}_{j,\mathbf{o}}| - w\,g_{j,\mathbf{o}}
\big\|_{L^2},
\end{equation}
where $\|\cdot\|_{L^2}$ denotes the Lebesgue $L^2$ norm~\cite{stein2009real}.
In practice, let $\mathbf{Y},\mathbf{G}\!\in\!\mathbb{R}^N$ be sampled magnitudes of 
$\hat{\psi}_{j,\mathbf{o}}$ and $g_{j,\mathbf{o}}$ within the ROI.
We compute $w_{j,\mathbf{o}}$ by ridge regression:
\begin{equation}
w_{j,\mathbf{o}}
=
\arg\min_{w}
\big\|
|\mathbf{Y}| - w\,\mathbf{G}
\big\|_2^2
+ \lambda w^2,
\label{eq:ridge_regression}
\end{equation}
where $\lambda$ is a small regularization term.
The resulting scalar is broadcast to four channels as $\mathbf{w}_{j,\mathbf{o}}$.
The pair $(g_{j,\mathbf{o}},\mathbf{w}_{j,\mathbf{o}})$ forms the canonical Gaussian. The complete transition bank is built as:
\begin{equation}
\big\{
(\mathbf{w}_{j,\mathbf{o}}, g_{j,\mathbf{o}}(\mathbf{x}))
\ \big|\ 
j,\mathbf{o}
\big\},
\label{eq:transition_bank}
\end{equation}
which maps each canonical wavelet kernel to its Gaussian counterpart:
$
\hat{\psi}_{j,\mathbf{o}} \rightarrow g_{j,\mathbf{o}}.
$
For $\hat{\psi}_{j,\mathbf{o},\mathbf{k}}$, the Gaussian approximation is retrieved by key $(j,\mathbf{o})$ and translated via
\begin{equation}
g_{j,\mathbf{o},\mathbf{k}}(\mathbf{x})
=
\big(T_{\,2^{j}\mathbf{k}+\boldsymbol{\delta}_{j,\mathbf{o}}}
g_{j,\mathbf{o}}\big)(\mathbf{x}),
\label{eq:gaussian_covariation}
\end{equation}
which directly parallels the translation property in Eq.\eqref{eq:covariation_law}.

\subsection{Analytical Gaussian Construction}
\label{sec:4.3}
We present an analytical formulation for directly representing a volumetric radiance field using Gaussian primitives. The process consists of four stages:
(i) The discrete wavelet transform (DWT) is first applied to decompose the radiance field into multiscale subbands. 
(ii) Each wavelet coefficient is then associated with a Gaussian primitive that models its spatial footprint. 
(iii) Leveraging the linearity of the inverse DWT (IDWT), the spatial contributions of all coefficients within each subband are analytically aggregated. 
(iv) All subband-level Gaussian fields are combined to yield a complete Gaussian representation of the original radiance field.

\textbf{Multichannel Wavelet Transform.}
The input volumetric radiance field is a four-channel function:
\begin{equation}
V : \mathbb{R}^3 \to \mathbb{R}^4,
\;\;
V(\mathbf{x}) =
\begin{bmatrix}
R(\mathbf{x}) \,\, G(\mathbf{x}) \,\, B(\mathbf{x}) \,\, \alpha(\mathbf{x})
\end{bmatrix}^{\!\top}.
\end{equation}
After performing a $J$-level 3D discrete wavelet transform (DWT), we obtain a set of coefficients:
\begin{equation}
\{A_{j,\mathbf{o},\mathbf{k},c} \in \mathbb{R}\,|\,j,\,\mathbf{o},\,k\}
,\;
c \in \{R,G,B,\alpha\},
\label{eq:mul_wavelet_coff}
\end{equation}
where $j$ denotes the decomposition level, $\mathbf{o}$ the subband orientation, and $\mathbf{k} = (k_x, k_y, k_z)$ the spatial index within the subband domain.
Each coefficient $A_{j,\mathbf{o},\mathbf{k},c}$ thus encodes the contribution of channel $c$ at spatial position $\mathbf{k}$ in the subband oriented along $\mathbf{o}$ at level $j$.

\textbf{Gaussian Construction from a Single Coefficient.}
Following Section~\ref{sec:4.2}, each subband $(j, \mathbf{o})$ is associated with a canonical Gaussian
$g(\mathbf{x};\;\mathbf{u}_{j, \mathbf{o}},\Sigma_{j, \mathbf{o}})$ (Eq.~\ref{eq:canonical_gaussian})
and a 4D modulation vector
$\mathbf{w}_{j, \mathbf{o}}\in\mathbb{R}^4$ (Eq.~\ref{eq:ridge_regression}).
For each coefficient at location $\mathbf{k}$,
we compute an RGBA modulation vector:
\begin{equation}
\mathbf{p}_{j, \mathbf{o},\mathbf{k}}
=
\begin{bmatrix}
|A_{j, \mathbf{o}, \mathbf{k}, R}|\\  
|A_{j, \mathbf{o}, \mathbf{k}, G}|\\
|A_{j, \mathbf{o}, \mathbf{k}, B}|\\
|A_{j, \mathbf{o}, \mathbf{k}, \alpha}|
\end{bmatrix}
\odot
\mathbf{w}_{j, \mathbf{o}},
\label{eq:per_location_weight}
\end{equation}
where $\odot$ denotes element-wise multiplication.
This vector encodes the local spectral energy of the RGBA channels, modulated by the channel weights.
Multiplying the RGBA vector $\mathbf{p}_{j,\mathbf{o},\mathbf{k}}$ with the geometric Gaussian
$g(\mathbf{x};, \mathbf{u}_{j,\mathbf{o}},, \Sigma_{j,\mathbf{o}})$ produces a 4D Gaussian primitive that represents the contribution of this wavelet coefficient:
\begin{equation}
s_{j,\mathbf{o}}\,
\mathbf{p}_{j,\mathbf{o}, \mathbf{k}}\,
g\!\left(\mathbf{x};\, \mathbf{u}_{j,\mathbf{o},\mathbf{k}},\, \Sigma_{j, \mathbf{o}}\right).
\label{eq:wavelet_gaussian_kernel}
\end{equation}
Here, $s_{j,\mathbf{o}}=2^{-3j/2}$ compensates for the intrinsic decay of the wavelet energy at coarser scales, while $\mathbf{u}_{j,\mathbf{o},\mathbf{k}}$ denotes the translated center of the canonical Gaussian, consistent with the canonical wavelet formulation in Eq.~\eqref{eq:covariation_law}:
\begin{equation}
\mathbf{u}_{j,\mathbf{o},\mathbf{k}}
= T_{2^j\mathbf{k}+\boldsymbol{\delta}_{j, \mathbf{o}}}(\mathbf{u}_{j,\mathbf{o}}).
\label{eq:translation_operator}
\end{equation}
In summary, each wavelet coefficient gives rise to a Gaussian primitive parameterized as
$
\big(
s_{j,\mathbf{o}}\,
\mathbf{p}_{j,\mathbf{o}, \mathbf{k}},
\;
\mathbf{u}_{j,\mathbf{o},\mathbf{k}},
\;
\Sigma_{j, \mathbf{o}}
\big),
$
where the RGBA amplitude corresponds to $s_{j,\mathbf{o}}\mathbf{p}_{j,\mathbf{o},\mathbf{k}}$,
the center to $\mathbf{u}_{j,\mathbf{o},\mathbf{k}}$,
and the covariance to $\Sigma_{j,\mathbf{o}}$.

\textbf{Gaussian Construction from a Subband.}
To construct the Gaussian representation of an entire subband $(j, \mathbf{o})$,
we first apply an adaptive filtering strategy (see supplementary material) to retain a sparse set of significant coefficients.
Let $\Omega_{j,\mathbf{o}}$ denote the set of retained indices;
each $\mathbf{k}\in\Omega_{j,\mathbf{o}}$ corresponds to a nonzero coefficient within the subband.
The sparsified subband can be written as a linear combination of impulses (Eq.~\eqref{eq:wavelet_impulse}):
$
\sum_{\mathbf{k}\in\Omega_{j,\mathbf{o}}}
A_{j,\mathbf{o},\mathbf{k},c}\,
e_{j,\mathbf{o},\mathbf{k}}(\boldsymbol{\kappa}).
$
Applying the IDWT to this expression yields the subband’s spatial response:
\begin{equation}
\mathcal{W}^{-1}\!\Big[
\sum_{\mathbf{k}\in\Omega_{j,\mathbf{o}}}
A_{j,\mathbf{o},\mathbf{k},c}\,
e_{j,\mathbf{o},\mathbf{k}}(\boldsymbol{\kappa})
\Big],\quad
c\!\in\!\{r,g,b,\alpha\}.
\label{eq:sparse_inverse_dwt}
\end{equation}
By linearity of $\mathcal{W}^{-1}$, Eq.~\eqref{eq:sparse_inverse_dwt} can be rewritten as:
\begin{align}
\sum_{\mathbf{k}\in\Omega_{j,\mathbf{o}}}
A_{j,\mathbf{o},\mathbf{k},c}\,
\mathcal{W}^{-1}\!\left[e_{j,\mathbf{o},\mathbf{k}}(\boldsymbol{\kappa})\right].
\end{align}
Substituting the inverse kernel (Eq.~\ref{eq:idwt_impluse}) gives:
\begin{equation}
\sum_{\mathbf{k}\in\Omega_{j,\mathbf{o}}}
A_{j,\mathbf{o},\mathbf{k},c}\,\hat{\psi}_{j,\mathbf{o},\mathbf{k}}(\mathbf{x}).
\end{equation}
Approximating $\hat{\psi}_{j,\mathbf{o},\mathbf{k}}(\mathbf{x})$ by its optimal Gaussian form (Eq.~\ref{eq:gaussian_covariation}) leads to:
\begin{equation}
\sum_{\mathbf{k}\in\Omega_{j,\mathbf{o}}}
A_{j,\mathbf{o},\mathbf{k},c}\,w_{j,\mathbf{o}}\,
g_{j,\mathbf{o},\mathbf{k}}(\mathbf{x}).
\end{equation}
After applying attenuation compensation $s_{j,\mathbf{o}}$ and jointly arranging the four RGBA channels, we obtain a linear combination of Gaussian primitives (Eq.~\eqref{eq:wavelet_gaussian_kernel}):
\begin{align}
\sum_{\mathbf{k}\in\Omega_{j,\mathbf{o}}}
s_{j,\mathbf{o}}\,
\mathbf{p}_{j,\mathbf{o},\mathbf{k}}\,
g\!\left(\mathbf{x};
\mathbf{u}_{j,\mathbf{o},\mathbf{k}},
\Sigma_{\mathbf{o},j}\right).
\label{eq:sparse_inverse_final}
\end{align}
Hence, the spatial response of the sparsified subband can be represented by a set of Gaussian primitives:
\begin{equation}
\big\{
s_{j,\mathbf{o}}\,
\mathbf{p}_{j,\mathbf{o},\mathbf{k}},
\;
\mathbf{u}_{j,\mathbf{o},\mathbf{k}},
\;
\Sigma_{j,\mathbf{o}}
\;\big|\;
\mathbf{k}\!\in\!\Omega_{j,\mathbf{o}}
\big\}.    
\end{equation}

\textbf{Gaussian Construction from Subbands Aggregation.}
By aggregating all subbands, we obtain a unified Gaussian-based volumetric representation that inherently preserves the coarse-to-fine information of the wavelet analysis:
\begin{equation}
RF(\mathbf{x})
\approx
\sum_{j=1}^{J}
\sum_{\mathbf{o}}
\sum_{\mathbf{k} \in \Omega_{j,\mathbf{o}}}
s_{j, \mathbf{o}}\,
\mathbf{p}_{j, \mathbf{o}, \mathbf{k}}\,
g\!\left(
\mathbf{x};\, \mathbf{u}_{j, \mathbf{o}, \mathbf{k}},\, \Sigma_{j, \mathbf{o}}
\right).
\label{eq:3dgs_reconstruction}
\end{equation}
Each term in Eq.~\eqref{eq:3dgs_reconstruction} serves a localized Gaussian primitive that contributes to the volumetric radiance field across scales and orientations.
The resulting GS set is defined as:
\begin{equation}
\mathcal{G} =
\Big\{
\big(
s_{j, \mathbf{o}}\,\mathbf{p}_{j, \mathbf{o}, \mathbf{k}},\;
\mathbf{u}_{j, \mathbf{o}, \mathbf{k}},\;
\Sigma_{j, \mathbf{o}}
\big)
\;\Big|\;
\mathbf{k} \in \Omega_{j,\mathbf{o}},\;
j,\mathbf{o}
\Big\},      
\end{equation}
where \( \mathbf{u}_{j, \mathbf{o}, \mathbf{k}},\,
\Sigma_{j, \mathbf{o}},\,s_{j, \mathbf{o}}\,\mathbf{p}_{j, \mathbf{o},\mathbf{k}}\) provide the center, covariance matrix, color and opacity for initializing the 3DGS representation.

\subsection{Image-space Gaussian Fine-tuning}
\label{sec:4.4}

To account for different visual modes in the volume, we generate multiple sets of radiance fields and their corresponding Gaussian representations using a series of transfer functions $\{ TF_h \}$, where each $TF_h$ is non-zero only within specific value intervals of the physical field $F$ and zero elsewhere. This results in a set of radiance fields $RF_h = TF_h(F)$ for each mode $h$.
For each radiance field $RF_h$, we render 64 images through volumetric integration, which provide supervision data for fine-tuning the initially constructed 3D Gaussian Set $\mathcal{G}_h$. The image-space loss is computed as in \cite{kerbl3Dgaussians} and backpropagated to update $\mathcal{G}_h$ . In practice, 5 transfer functions are selected for each volume. After fine-tuning, the Gaussians $\mathcal{G}_h$ for each mode $h$ can be either concatenated or separately rendered for global and local visibility, as described in \cite{tang2025ivrgs}.

%% file: sec/5_exp.tex
\begin{figure}[t]
    \centering
    \includegraphics[width=1.0\linewidth]{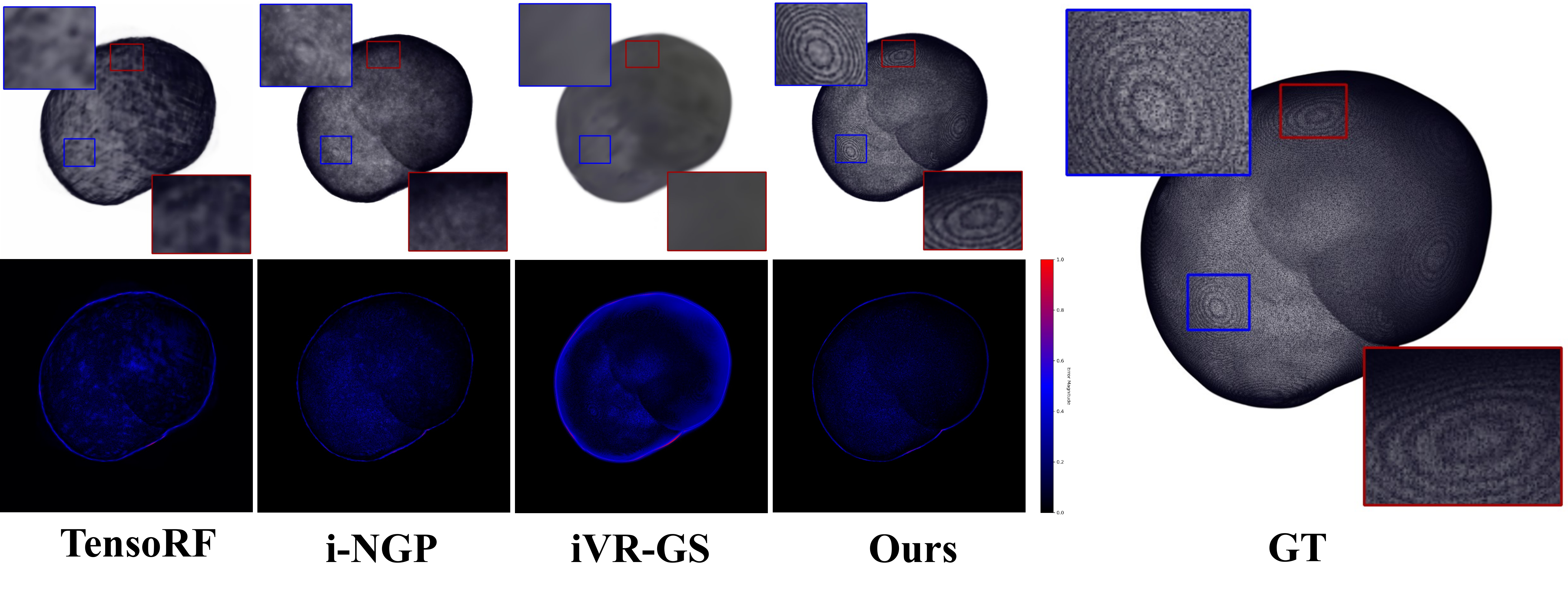}
    \caption{
    Qualitative comparison at an early training stage (0.5K). Our VBM-based Gaussians already captures fine structural details, demonstrating strong initialization and efficient convergence.
    }
    \label{fig:detail}
\end{figure}

%\begin{figure}[t]
%    \centering
%    \includegraphics[width=1\linewidth]{fig/psnr_comparison_siggraph.pdf}
%    \caption{
%    PSNR convergence curve on the Supernova dataset, showing that the VBM-based Gaussian closely tracks the optimal solution and achieves fast and stable convergence.}
%    \label{fig:psnr_comparison_siggraph}
%\end{figure}

\section{Experiments}
\label{sec:exp}

\begin{figure*}[t]
    \centering
    \includegraphics[width=1.0\linewidth]{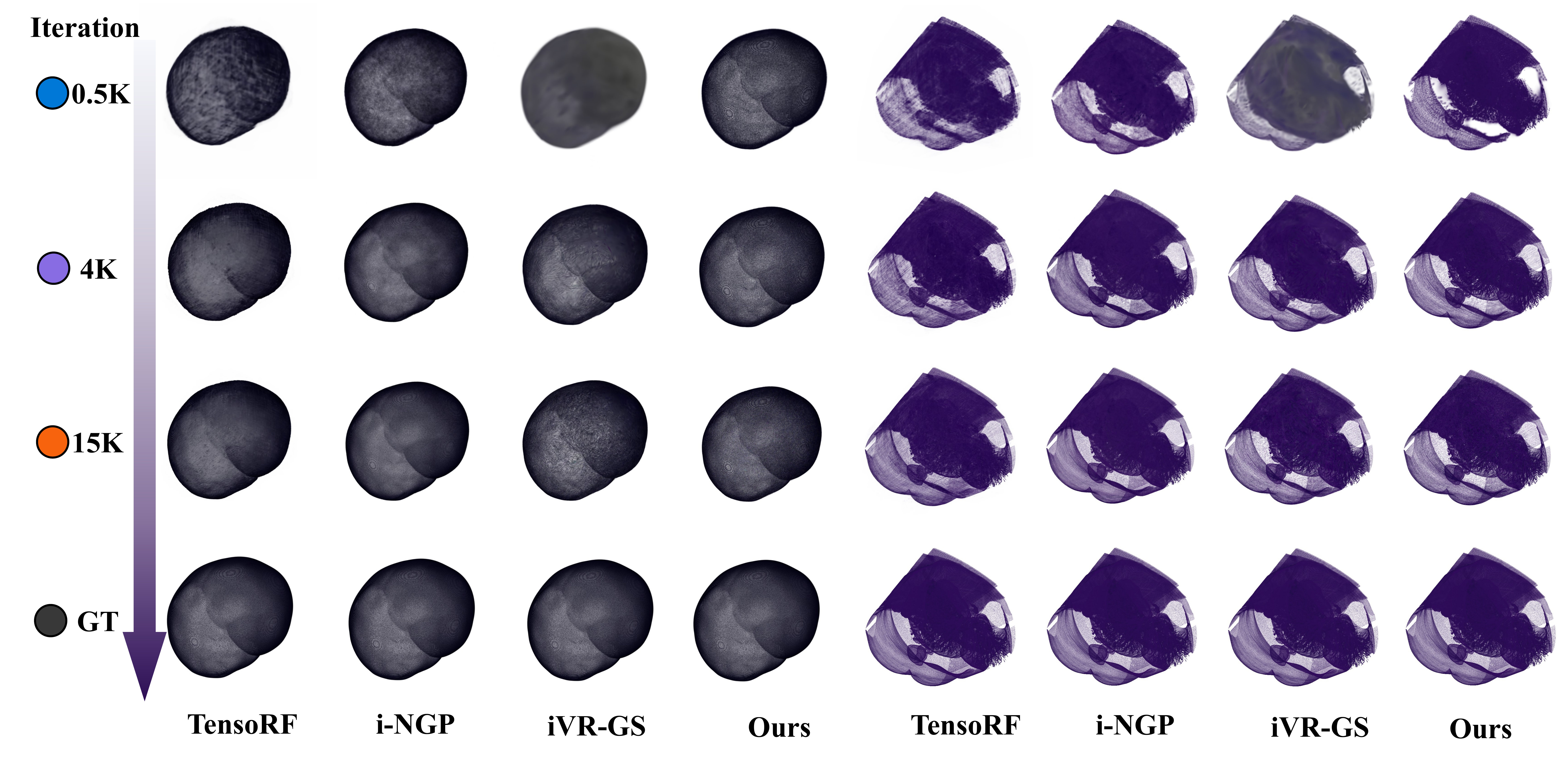}
      \caption{
    Qualitative comparison across training iterations. Our method rapidly refines structural details on both the Supernova simulation (left) and the Colon Prone CT volume (right), demonstrating consistent convergence across scientific and medical data.
    }
    \label{fig:render-comparision-1}
\end{figure*}

\begin{table}[t]
\centering
\small
\caption{
Qualitative comparison on four datasets. PSNR/SSIM are reported at 4k and 15k iterations. Bold text indicates the best results, and underlined text represents the second-best.}
\renewcommand{\arraystretch}{1.15}
\setlength{\tabcolsep}{5pt}
\begin{tabular}{l|cc|cc|c}
\hline
\multirow{2}{*}{\textbf{Method}} &
\multicolumn{2}{c|}{\textbf{4k Iter}} &
\multicolumn{2}{c|}{\textbf{15k Iter}} &
\multirow{2}{*}{\textbf{Time (min)}} \\ \cline{2-5}
 & \textbf{PSNR} & \textbf{SSIM} & \textbf{PSNR} & \textbf{SSIM} & \\ \hline

% ================= Supernova ==================
\multicolumn{6}{c}{\textit{\textbf{Supernova}}} \\ \hline
TensoRF      & 32.78 & 0.9022 & 34.01 & 0.9100 & 7.58  \\
InstantNGP   & 33.43 & 0.9176 & 35.40 & 0.9207 & \underline{5.17}  \\
3DGS         & \underline{38.20} & \underline{0.9258} & 39.11 & 0.9231 & 5.73  \\
iVR-GS       & 38.02 & 0.9243 & \underline{39.23} & \underline{0.9246} & 6.18  \\
\textbf{Ours}& \textbf{39.09} & \textbf{0.9307} & \textbf{40.93} & \textbf{0.9273} & \textbf{5.06}  \\ \hline

% ================= Colon Prone ==================
\multicolumn{6}{c}{\textit{\textbf{Colon Prone}}} \\ \hline
TensoRF      & 26.40 & 0.8860 & 28.08  & 0.8859 & 12.65 \\
InstantNGP   & \underline{27.81} & 0.9109 & \textbf{29.75} & \textbf{0.9245} & \textbf{5.85}  \\
3DGS         & 26.65 & \underline{0.9123} & 28.89 & 0.9192 & 7.93  \\
iVR-GS       & 26.23 & 0.9105 & 28.91 & 0.9182 & 8.16  \\
\textbf{Ours}& \textbf{28.23} & \textbf{0.9208} & \underline{29.69} & \underline{0.9198} & \underline{7.86}  \\ \hline

% ================= Skull ==================
\multicolumn{6}{c}{\textit{\textbf{Skull}}} \\ \hline
TensoRF      & 25.97 & 0.925 & \textbf{28.00} & \underline{0.9460} & 14.20  \\
InstantNGP   & \underline{26.90} & 0.9313 & \underline{27.90} & \textbf{0.9488} & \textbf{4.70}  \\
3DGS         & 26.34 & 0.9372 & 27.50 & 0.9452 & 6.41  \\
iVR-GS       & 26.35 & \underline{0.9376} & 27.34 & 0.9446 & 7.45  \\
\textbf{Ours}& \textbf{27.18} & \textbf{0.9464} & 27.75 & 0.9454 & \underline{6.02}  \\ \hline

% ================= Foot ==================
\multicolumn{6}{c}{\textit{\textbf{Foot}}} \\ \hline
TensoRF      & \textbf{28.55} & 0.9219 & \textbf{30.75} & \textbf{0.9359} & 18.69  \\
InstantNGP   & 27.26 & \underline{0.9294} & 28.21 & \underline{0.9319} & \underline{5.87}  \\
3DGS         & 27.40 & 0.9328 & 28.01 & 0.9305 & 6.01  \\
iVR-GS       & 27.38 & 0.9327 & 27.98 & 0.9292 & 7.92  \\
\textbf{Ours}& \underline{27.98} & \textbf{0.9351} & \underline{28.23} & 0.9313 & \textbf{5.85}  \\ 
\hline
\end{tabular}
\label{tab:exp_results}
\end{table}

\subsection{Datasets and Metrics}
We evaluate our method on several representative volumetric datasets commonly used in scientific visualization. 
These include the \emph{Supernova} volume (a $432^3$ floating-point simulation from ECNR~\cite{ECNR}), 
as well as three medical datasets from~\cite{klacansky2017openscivis}:
the \emph{Colon Prone} CT scan ($512\times512\times463$, 16-bit intensity data), and two X-ray datasets, \emph{Skull} and \emph{Foot} , both of size $256^3$ with 8-bit intensity values. These datasets span diverse modalities and resolutions, ranging from dense physical simulations to medical scans with varying dynamic ranges. For quantitative evaluation, we employ two standard image-quality metrics: PSNR and SSIM.

\subsection{Implementation Details}
Our framework is implemented in PyTorch with CUDA acceleration and uses PyWavelets for multi-scale decomposition. The Adam optimizer is adopted for all experiments. Reference volume renderings are generated in ParaView 5.12.1 with NVIDIA IndeX serving as the baseline renderer. All experiments are conducted on a workstation equipped with NVIDIA RTX 3090 GPUs.

\subsection{Comparisons and Evaluations}
We compare our method against state-of-the-art neural scene representation approaches, including TensoRF~\cite{chen2022tensorf}, Instant-NGP~\cite{muller2022instant}, 3DGS~\cite{kerbl3Dgaussians}, and iVR-GS~\cite{tang2025ivrgs}. As summarized in Table~\ref{tab:exp_results}, which presents the average metrics for each scene using 5 transfer functions $TF_h$, our framework consistently achieves higher PSNR and SSIM across all datasets while requiring significantly fewer training iterations. Notably, even at an early training stage (4k iterations), our model already produces comparable result to those of existing methods after full convergence. This demonstrates the fast convergence, representational efficiency, and stable optimization behavior of our formulation. 

Qualitative comparisons are presented in Fig.~\ref{fig:detail} and Fig.~\ref{fig:render-comparision-1}, where our method yields sharper boundaries and more faithful volumetric structures than competing approaches, validating the effectiveness of the proposed wavelet-to-Gaussian mapping. For the \emph{Supernova} dataset, our model utilizes approximately \(1.8\times10^5\) Gaussian primitives and achieves real-time rendering performance of about 120~FPS on an RTX~3090 GPU.

\begin{figure}[t]
    \centering
    \includegraphics[width=1.0\linewidth]{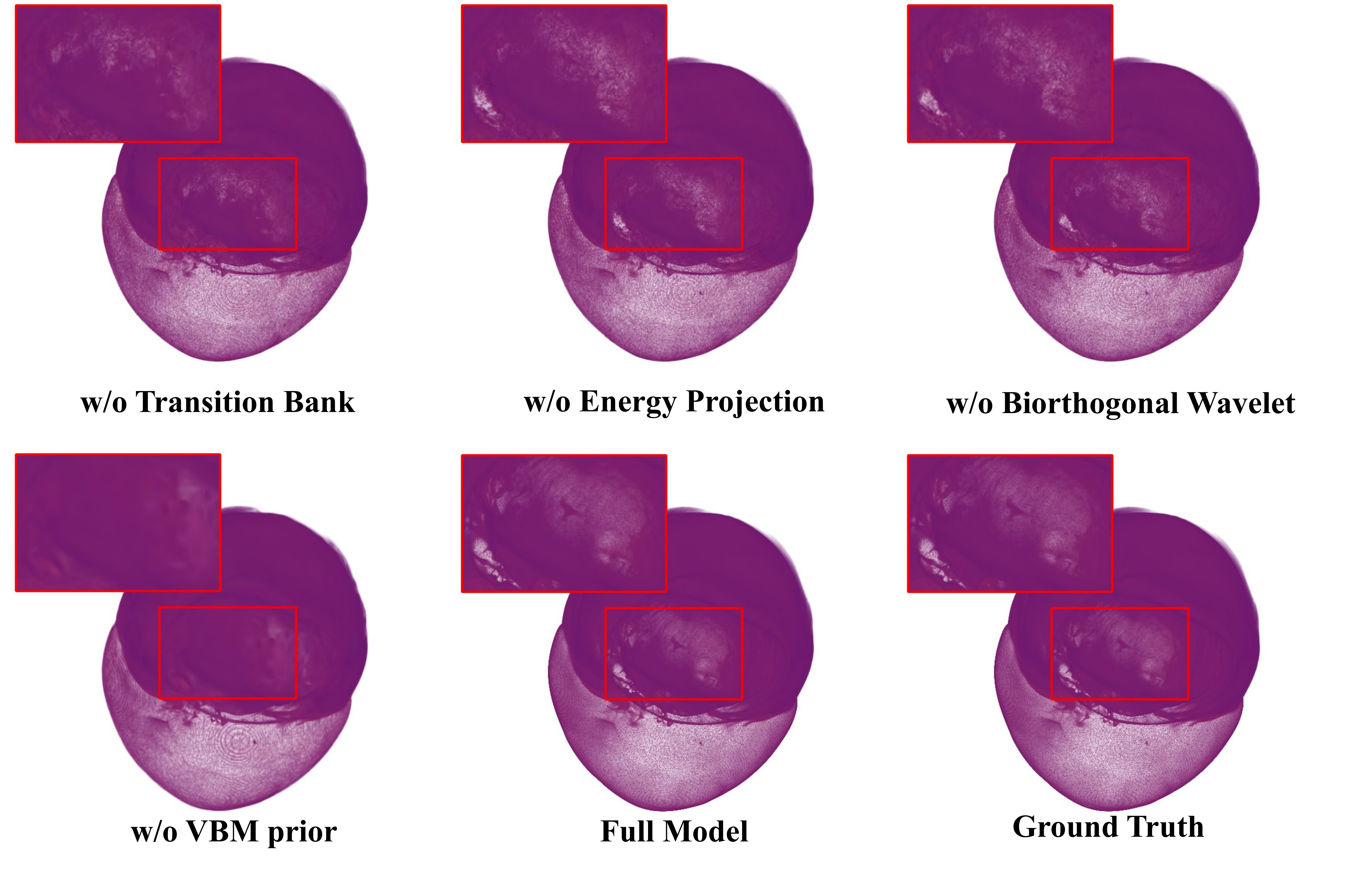}
    \caption{Qualitative ablation results of different variants on the \textit{Supernova} dataset using a selected pink TF.}
    \label{fig:ablation}
\end{figure}

\subsection{Ablation Studies}
We further analyze the contribution of each key component using the \emph{Supernova} dataset under identical training conditions. 
Quantitative results are reported in Table~\ref{tab:ablation}, and qualitative comparisons are presented in Fig.~\ref{fig:ablation}.

\textit{(1) w/o Transition Bank.} 
We disable the Wavelet-to-Gaussian Transition Bank by collapsing all covariances into a single shared isotropic matrix. 
Centers and opacity/appearance parameters remain unchanged, while each covariance is fixed as $\Sigma_{j,\mathbf{o}} = \sigma_{\text{iso}}^2 \mathbf{I}$ for all scales $j$ and orientations $\mathbf{o}$, where $\sigma_{\text{iso}}$ corresponds to the average spatial extent in the full model. 
Removing scale–orientation adaptivity produces smoother but noticeably blurrier results.

\textit{(2) w/o Energy Projection.} 
Eliminating the energy conversion (by setting $\mathbf{w}_{j,\mathbf{o}}\!=\!1$) causes global intensity imbalance, highlighting the importance of proper energy scaling for stable photometric consistency.

\textit{(3) w/o Biorthogonal Wavelet.} 
To evaluate sensitivity to the wavelet basis, we replace the default biorthogonal wavelet (\texttt{bior4.4}) with the Haar. Haar lacks the approximate translation consistency, leading to misaligned Gaussian centers and mild spatial fuzziness in the outputs.

\textit{(4) w/o VBM Initialization.} 
We eliminate the VBM-based Gaussian initialization and randomly set all Gaussian parameters. This results in suboptimal optimization performance, highlighting the importance of the proposed framework in providing a crucial prior for efficient and stable training.

\begin{table}[t]
\small
\centering
\caption{Ablation results on the \textit{Supernova} dataset, with PSNR (dB) and SSIM for quality evaluation.}

\label{tab:ablation}
\begin{tabular}{lcc}
\toprule
Variant & PSNR & SSIM \\
\midrule
 w/o Transition Bank & 30.88  & 0.9483  \\
 w/o Biorthogonal Wavelet & 31.94 & 0.9505 \\
 w/o Energy Projection & 32.18 & 0.9521 \\
 w/o VBM prior & 30.84 & 0.9469  \\
 Full model & \textbf{32.56} & \textbf{0.9543} \\
\bottomrule
\end{tabular}
\end{table}

\section{Conclusion}
We introduced \emph{VBM}, a unified framework that bridges volumetric wavelet analysis with 3D Gaussian Splatting for real-time volumetric visualization. Unlike heuristic or image-driven initialization strategies used in prior 3DGS-based methods, our approach establishes a principled analytical mapping from wavelet kernels to Gaussian primitives. By precomputing a compact Wavelet-to-Gaussian Transition Bank and formulating an analytical Gaussian construction strategy, VBM enables fast, accurate, and stable field-to-primitive conversion. A lightweight image-space fine-tuning further enhances visual fidelity without compromising efficiency. Extensive experiments demonstrate that VBM not only accelerates convergence and improves rendering quality but also preserves volumetric semantics during conversion, achieving both analytical rigor and practical efficiency. Beyond these performance gains, our framework provides a new mathematical foundation that unifies multiresolution analysis with explicit scene representations. We believe this paradigm will inspire future research in structured neural representations, scientific visualization, and real-time rendering of complex volumetric phenomena.

%% file: sec/X_supp.tex
The supplementary materials include a detailed presentation of the adaptive filtering strategy, along with a concise proof of translation consistency. For a more abstract and general theoretical treatment, we refer the reader to~\cite{antoine1999wavelets}.

\section{Adaptive Filter Strategy}

We describe the sparsification strategy employed in our wavelet-domain filtering pipeline. 
The approach integrates (1) robust noise estimation,  
(2) energy-aware cross-band allocation, and  
(3) multi-channel coherent coefficient selection.

\subsection{Global Budget Allocation Across Subbands}

Let $\{A_{j,\mathbf{o},\mathbf{k},c}\}$ denote the discrete wavelet coefficients of a multichannel signal,
indexed by scale $j$, orientation $\mathbf{o}$, spatial index $\mathbf{k}\!\in\!\mathbb{Z}^3$,
and channel $c \in \{R,G,B,\alpha\}$.
Let $c_{\mathrm{ref}}$ be the reference channel (typically $\alpha$). For each subband $(j,\mathbf{o})$ we define:

\begin{equation}
E_{j,\mathbf{o}} =
\sum_{\mathbf{k}}
\bigl| A_{j,\mathbf{o},\mathbf{k},c_{\mathrm{ref}}} \bigr|^2,
\end{equation}

\begin{equation}
N_{j,\mathbf{o}} = \#\{ \mathbf{k} \}.
\end{equation}
where $\#$ denotes the spatial cardinality of the subband, i.e., the total number of coefficient locations in the subbband $(j,\mathbf{o})$. The global sparsity budget $K_{\mathrm{tot}}$ is distributed according to:

\begin{equation}
S_{j,\mathbf{o}} = E_{j,\mathbf{o}}^{\alpha} \, N_{j,\mathbf{o}}^{\beta}.
\end{equation}
The assigned coefficient count is:

\begin{equation}
K_{j,\mathbf{o}} = 
\max \left( 10,\ 
\left\lfloor
K_{\mathrm{tot}}\, \frac{S_{j,\mathbf{o}}}{\sum_{j,\mathbf{o}} S_{j,\mathbf{o}}}
\right\rfloor
\right).
\end{equation}
Thus the allocation satisfies:

\begin{equation}
\sum_{j,\mathbf{o}} K_{j,\mathbf{o}} \le K_{\mathrm{tot}}.
\end{equation}
This scheme resembles rate–distortion optimized bit allocation,
but acts directly on coefficient cardinality rather than entropy.

\subsection{Joint Multi-Channel Sparsification}

For each subband we group coefficients across channels:

\begin{equation}
\mathbf{A}_{j,\mathbf{o},\mathbf{k}}
=
\bigl(
A_{j,\mathbf{o},\mathbf{k},R},\,
A_{j,\mathbf{o},\mathbf{k},G},\,
A_{j,\mathbf{o},\mathbf{k},B},\,
A_{j,\mathbf{o},\mathbf{k},\alpha}
\bigr)\in\mathbb{R}^4.
\end{equation}
Define the joint magnitude:

\begin{equation}
v_{j,\mathbf{o}}(\mathbf{k}) 
= \Vert \mathbf{A}_{j,\mathbf{o},\mathbf{k}} \Vert_2
= \sqrt{
\sum_{c}
\left| A_{j,\mathbf{o},\mathbf{k},c} \right|^2
}.
\end{equation}
Coefficient selection is performed on this joint norm,
ensuring that a spatial location is either retained or removed
\emph{coherently across channels}.

\subsection{Robust MAD-Based Thresholding}

We estimate a robust noise scale using the Median Absolute Deviation:
\begin{align}
\mathrm{MAD} &= 
\operatorname{median}_{\mathbf{k}}
\left| v_{j,\mathbf{o}}(\mathbf{k}) -
\operatorname{median}(v_{j,\mathbf{o}}) \right|, \\
\hat{\sigma} &= \frac{\mathrm{MAD}}{0.6745}.
\end{align}
Given a multiplier $\lambda > 0$, the subband threshold is:
\begin{equation}
T = \lambda \hat{\sigma}.
\end{equation}
We first define a preliminary mask:
\begin{equation}
M_{\mathrm{th}}(\mathbf{k}) 
= \mathbf{1}\bigl( v_{j,\mathbf{o}}(\mathbf{k}) \ge T \bigr).
\end{equation}

\subsection{Top-\texorpdfstring{$K$}{K} Selection Under Sparsity Constraints}

Let$\mathcal{C} = \{ \mathbf{k} \mid M_{\mathrm{th}}(\mathbf{k}) = 1 \}\;,v_i = v_{j,\mathbf{o}}(\mathbf{k}_i).$
We retain:
\begin{equation}
K = \min(K_{j,\mathbf{o}},\, |\mathcal{C}|).
\end{equation}
Let $v_{(K)}$ denote the $K$-th largest value in $\{v_i\}$.
The final selection mask is:

\begin{equation}
M(\mathbf{k}) = 
\mathbf{1}\Bigl( 
v_{j,\mathbf{o}}(\mathbf{k}) \ge \max(T,\, v_{(K)})
\Bigr).
\end{equation}
This enforces both a robust MAD-based noise floor $T$, and  
an exact cardinality constraint $K_{j,\mathbf{o}}$.

Unlike convex $\ell_1$ shrinkage,
this yields a strictly $\ell_0$-bounded representation.

\subsection{Sparse Coordinate Encoding}

Define the retained index set:

\begin{equation}
\mathcal{I}_{j,\mathbf{o}}
=
\{ \mathbf{k} \mid M(\mathbf{k}) = 1 \}.
\end{equation}

The sparse representation consists of:

\begin{equation}
\texttt{idx} = [\mathbf{k}] \in \mathbb{Z}^{K \times 3},
\qquad
\texttt{vals} = 
\bigl[
\widetilde{A}_{j,\mathbf{o},\mathbf{k},c}
\bigr] \in \mathbb{R}^{K}.
\end{equation}

This form is channel-coherent, compact,
and suitable for storage, entropy coding, or learned decoding.

\section{Proof of the Transition Consistency of the Wavelet Kernels}
We now show that within any fixed wavelet subband, the spatial response of an impulse located at an arbitrary coefficient index is exactly a translated copy of the canonical kernel. This follows from the multirate factorization of the inverse wavelet transform and two fundamental operator identities: upsampling--translation commutation and convolution--translation commutation.

\begin{theorem}[Translation Consistency Within a Wavelet Subband]
Let $\mathcal{W}^{-1}$ denote the inverse 3D discrete wavelet transform implemented by a multirate, separable, finite-support synthesis filter bank with $J$ decomposition levels.
For each subband $(j,\mathbf{o})$, let
\[
H_{j,\mathbf{o}} : \ell^2(\mathbb{Z}^3) \to \ell^2(\mathbb{Z}^3)
\]
denote the partial reconstruction operator mapping coefficients in that subband back to the spatial domain.
Define the impulse wavelet responses:
\[
\hat{\psi}_{j,\mathbf{o}} := H_{j,\mathbf{o}} \delta_{\mathbf{0}}, \qquad
\hat{\psi}_{j,\mathbf{o},\mathbf{k}} := H_{j,\mathbf{o}} \delta_{\mathbf{k}}.
\]
Then for every $\mathbf{k} \in \mathbb{Z}^3$,
\[
\hat{\psi}_{j,\mathbf{o},\mathbf{k}} = T_{2^j \mathbf{k}} \, \hat{\psi}_{j,\mathbf{o}}
\]
where $T_{\mathbf{n}}$ denotes the discrete translation operator
\[
(T_{\mathbf{n}} f)(\mathbf{x}) = f(\mathbf{x} - \mathbf{n}).
\]
\end{theorem}

\begin{proof}
We work entirely in $\ell^2(\mathbb{Z}^3)$, which denotes the Hilbert space of square-summable
3D discrete signals:
\[
\ell^2(\mathbb{Z}^3)
=
\left\{ f : \mathbb{Z}^3 \to \mathbb{R}
\ \middle| \ 
\sum_{\mathbf{x}\in \mathbb{Z}^3} |f(\mathbf{x})|^2 < \infty
\right\},
\]
All signals are discrete; no approximation is involved.

\vspace{0.5em}
\noindent\textbf{Step 1. Multirate structure of inverse DWT.}

For each subband $(j,\mathbf{o})$, the inverse wavelet transform factorizes as
\[
H_{j,\mathbf{o}} = C_{h_{j,\mathbf{o}}} \, U_j,
\]
where:
\begin{itemize}
\item $U_j$ is 3D upsampling by factor $2^j$:
  \[
  (U_j c)(\mathbf{x}) =
  \begin{cases}
  c(\mathbf{x}/2^j), & \mathbf{x} \equiv \mathbf{0} \pmod{2^j}, \\[2pt]
  0, & \text{otherwise},
  \end{cases}
  \]
\item $C_{h_{j,\mathbf{o}}}$ is convolution with a finite-support sequence:
  \[
  (C_{h} c)(\mathbf{x}) = \sum_{\mathbf{n}\in\mathbb{Z}^3} h(\mathbf{x}-\mathbf{n}) \, c(\mathbf{n}).
  \]
\end{itemize}
This is a standard identity from multirate filter bank theory.

\vspace{0.5em}
\noindent\textbf{Step 2. Two operator identities.}

\begin{lemma}[Upsampling + Translation]
For any $\mathbf{k} \in \mathbb{Z}^3$,
\[
U_j T_{\mathbf{k}} = T_{2^j \mathbf{k}} U_j.
\]
\end{lemma}
\begin{proof}
For any $c$ and $\mathbf{x}$,
\begin{align*}
(U_j T_{\mathbf{k}} c)(\mathbf{x})
&= \begin{cases}
(T_{\mathbf{k}} c)(\mathbf{x}/2^j), & \mathbf{x}\equiv0 \pmod{2^j}, \\
0, & \text{otherwise}
\end{cases} \\
&= \begin{cases}
c(\mathbf{x}/2^j - \mathbf{k}), & \mathbf{x}\equiv0 \pmod{2^j}, \\
0, & \text{otherwise}
\end{cases} \\
&= (U_j c)(\mathbf{x}-2^j\mathbf{k}) = (T_{2^j\mathbf{k}} U_j c)(\mathbf{x}).
\end{align*}
\end{proof}

\begin{lemma}[Convolution commutes with translation]
\[
C_h T_{\mathbf{n}} = T_{\mathbf{n}} C_h.
\]
\end{lemma}
\begin{proof}
Direct substitution:
\begin{align*}
(C_h T_{\mathbf{n}}c)(\mathbf{x})
&= \sum_{\mathbf{m}} h(\mathbf{x}-\mathbf{m}) \, c(\mathbf{m}-\mathbf{n}) \\
&= \sum_{\mathbf{p}} h((\mathbf{x}-\mathbf{n})-\mathbf{p}) \, c(\mathbf{p}) \\
&= (C_h c)(\mathbf{x}-\mathbf{n}) = (T_{\mathbf{n}} C_h c)(\mathbf{x}).
\end{align*}
\end{proof}

\vspace{0.5em}
\noindent\textbf{Step 3. Combine the lemmas.}
Using $H_{j,\mathbf{o}} = C_{h_{j,\mathbf{o}}} U_j$,
\begin{align*}
H_{j,\mathbf{o}} T_{\mathbf{k}}
&= C_{h_{j,\mathbf{o}}} (U_j T_{\mathbf{k}}) \\
&= C_{h_{j,\mathbf{o}}} (T_{2^j\mathbf{k}} U_j)
\quad \text{(Lemma 1)} \\
&= T_{2^j\mathbf{k}} (C_{h_{j,\mathbf{o}}} U_j)
\quad \text{(Lemma 2)} \\
&= T_{2^j\mathbf{k}} H_{j,\mathbf{o}}.
\end{align*}

\vspace{0.5em}
\noindent\textbf{Step 4. Apply to impulses.}
Since
\[
\delta_{\mathbf{k}} = T_{\mathbf{k}} \delta_{\mathbf{0}},
\]
we have
\begin{align*}
\hat{\psi}_{j,\mathbf{o},\mathbf{k}}
&= H_{j,\mathbf{o}} \delta_{\mathbf{k}}
= H_{j,\mathbf{o}} T_{\mathbf{k}} \delta_{\mathbf{0}} \\
&= T_{2^j\mathbf{k}} H_{j,\mathbf{o}} \delta_{\mathbf{0}}
= T_{2^j\mathbf{k}} \hat{\psi}_{j,\mathbf{o}}.
\end{align*}
This completes the proof.
\end{proof}

\textbf{Remark.} If the synthesis filters are \textbf{linear-phase}, then the effective synthesis kernel satisfies
\[
h_{j,\mathbf{o}}(\mathbf{n}) = \tilde{h}_{j,\mathbf{o}}(\mathbf{n}-\boldsymbol{\delta}_{j,\mathbf{o}}),
\]
with fixed $\boldsymbol{\delta}_{j,\mathbf{o}} \in \mathbb{R}^3$ depending only on the filter bank. Embedding the discrete kernel into $\mathbb{R}^3$, one obtains:
\[
\hat{\psi}_{j,\mathbf{o},\mathbf{k}}(\mathbf{x})
\approx
\hat{\psi}_{j,\mathbf{o}}(\mathbf{x} - (2^j \mathbf{k} + \boldsymbol{\delta}_{j,\mathbf{o}})),
\]
where the ``$\approx$'' arises only from the discrete-to-continuous interpolation. But on the discrete grid, the result above is exact.

\begin{figure*}[]
    \centering
    \includegraphics[width=1.0\linewidth]{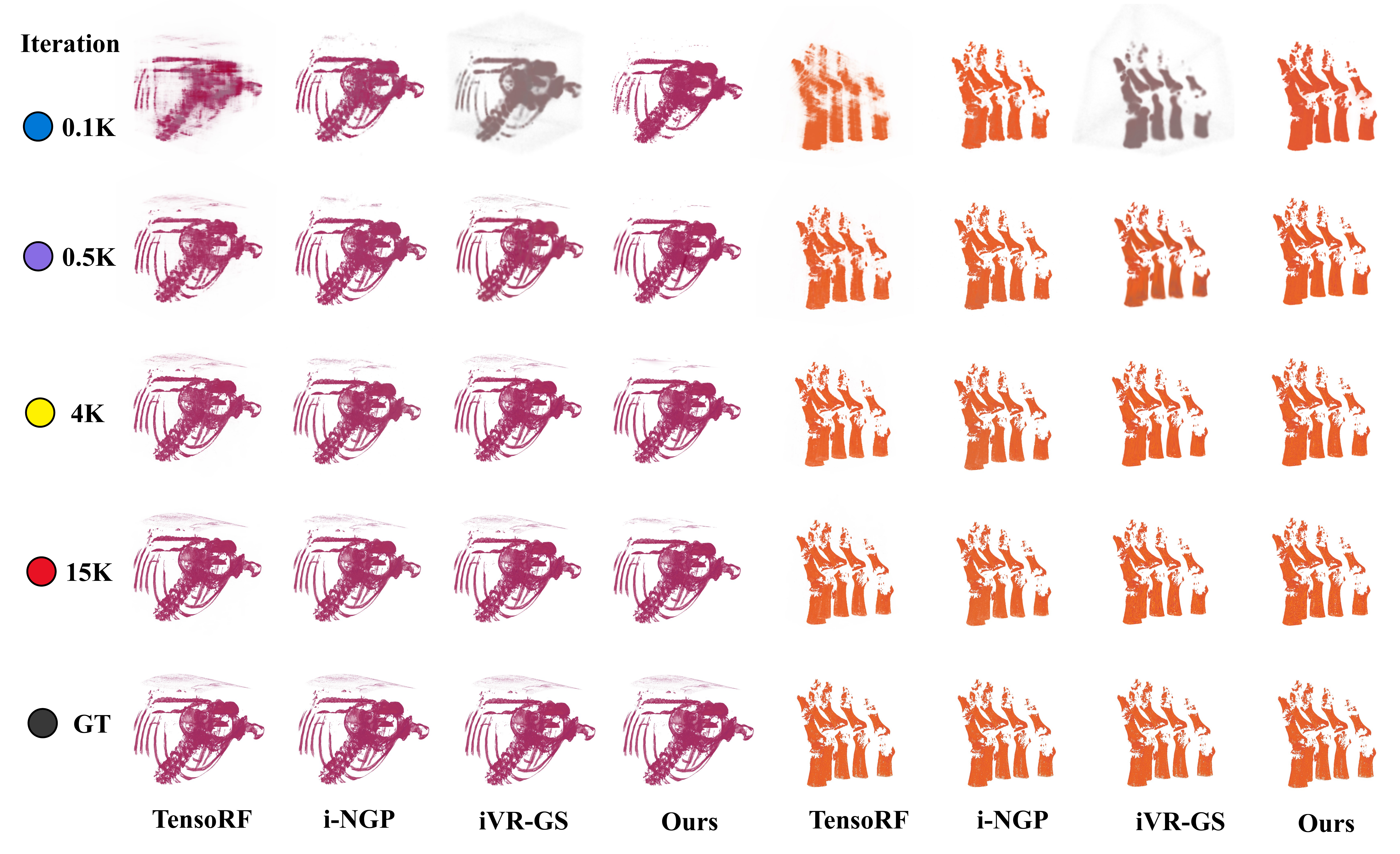}
    \caption{Qualitative comparison across training iterations on the \emph{Skull} and \emph{Foot} dataset. }
    \label{fig:render-comparision-2}
\end{figure*}